# Development of current estimated household data and agent-based simulation of the future population distribution of households in Japan

Kajiwara Kento[1], Jue Ma[2], Toshikazu Seto[3], Yoshihide Sekimoto[4], Yoshiki Ogawa[4], Hiroshi Omata[4]

[1] Mitsui Fudosan Co., Ltd, Tokyo, Japan
[2] School of Engineering, The University of Tokyo, Tokyo, Japan
[3] Department of Geography, Komazawa University, Tokyo, Japan
[4] Center for Spatial Information Science, The University of Tokyo, Tokyo, Japan

*Corresponding author. Email: majue@iis.u-tokyo.ac.jp; Address: Ce509, 4-6-1, Komaba, Merugo-ku, Tokyo 153-8505, Japan

**Acknowledgements**

This research did not receive any specific grand from funding agencies in the public, commercial, or not-for-profit sectors. And this work was supported by the Nanto city office of Japan and JoRAS by Center of Spatial Information Science, The University of Tokyo.

**Declarations of interest: None**

# Development of current estimated household data and agent-based simulation of the future population distribution of households in Japan

**Abstract**

In response to the declining population and aging infrastructure in Japan, local governments are implementing compact city policies such as the location normalization plan. To optimize the reorganization of urban public infrastructure, it is important to provide detailed and accurate forecasts of the distribution of urban populations and households. However, many local governments do not have the necessary data and forecasting capability. Moreover, current forecasts of gender- and age-based population data only exist at the municipal level, and household data are only available by family type at the prefecture level. Meanwhile, the accuracy is limited with an assumption of same change rate of population in all municipalities and within each city. Therefore, the aim of this study was to develop an agent-based microsimulation household transition model, with the household as the unit and agent, and household data was estimated for all cities in Japan from 2015. Estimated household data comprised the family type, house type, and address, age, and gender of household members, obtained from the national census, and building data. The resulting household transition model was used to forecast the attributes of each household every five years. Simulations in Toyama and Shizuoka Prefectures, Japan from 1980 to 2010 provided highly accurate estimates of municipal-level population by age and household volume by family type. The proposed model was also applied to predict the future distribution of disappearing villages and vacant houses in Japan.

**Keywords**

population forecast, household, microsimulation model, vacant house, disappearing village

1. **Introduction**

In May 2014, the Masuda Report concluded that half of all municipalities in Japan will likely disappear by 2040, with the population of rural areas in particular predicted to decline as a result of the decreasing birthrate, aging population, and population concentration in the Tokyo metropolitan area (Masuda, 2014). In addition, local governments are implementing compact city policies to combat aging infrastructure built predominantly during the period of rapid economic growth in the 1955-1973 (Miyauchi, Setoguchi & Ito, 2021; Hattori, Kaido & Matsuyuki, 2017). Specifically, the Ministry of Land, Infrastructure, Transport and Tourism (MLIT) has developed a policy based on location normalization plans, with 559 out of 1,724 municipalities in Japan preparing plans of compact city for future urban restructuring as of December 2020 (Ministry of Land, Infrastructure, Transport and Tourism, 2021). To facilitate the reorganization of schools, public facilities, public transportation systems, and designated residential areas, it is vital to accurately estimate future population and household data (Karashima, Ohgai & Saito, 2014; Matsunaka et al., 2013; Tanura, Iwamoto & Tanaka, 2018; Tsuboi, Ikaruga & Kobayashi, 2016).

However, detailed future population and household estimates are currently unavailable in Japan. Such data have been incorporated into various urban models, but only for major cities or at the municipal level (Akkerman & Shimoura, 2012; Abe, 2011; Lee et al., 2021). Many local governments use estimated future population and household data provided by the National Institute of Population and Social Security Research (NIPSSR) for urban planning; however, NIPSSR is based on the 2015 national census, and provides detailed forecasts of population by gender and five-year age group only at the municipality level, as well as household data by family type only at the prefecture level (National Institute of Population and Social Security Research, 2018). MLIT then estimates the future population of Japan using a 500-m mesh according to NIPSSR data; however, the accuracy is limited because the estimation assumes the same change rate in all municipalities and within each city, without considering the concentration of populations around stations or central areas (National Land Policy Bureau, MLIT, 2015).

A number of studies have attempted to forecast future population and household numbers. In Japan, for example, NIPSSR used the cohort factor method[1] for estimating population and the householder rate method[2] for estimating households (National Institute of Population and Social Security Research, 2018). Moreover, the urban planning support system MyCityForecast was developed to simulate planning policy results for collaboration between citizens and governments (Hasegawa et al., 2019). This system estimates the population in 500-m-mesh units by referring to the NIPSSR method at the municipal level. Additionally, several previous studies have been performed for population estimation. Such as population distribution transition modeling for examining the effectiveness of compact city policy (Sugimoto et al., 2018), microsimulation modeling for population estimation based on the location choice model (Keito & Hao, 2019), household simulations based on a mesoscopic model (Yamagiwa et al., 2017), population distribution estimates to determine the effect of measures designed to attract residents (Chikuma & Sato, 2017), and population change estimation according to family type (Ishigami & Kurokawa, 2001). However, these methods are applicable only to parts of municipalities or limited areas, rather than the whole country with the limitation of data availability.

On the other hand, future population and household forecast are being developed worldwide (Christiansen & Keilman, 2013; Keilman, 2016; Murphy, 1991; Wilson, 2013; Zhuge et al., 2018). Münnich et al. (2021) constructed a national microsimulation model for Germany called MikroSim, accounting for local variations in each of communities. Ballas et al. (2005) dynamically simulate the entire population of Britain at the small area level by a spatial microsimulation model named as SimBritain. Hecht et al. (2018) realized the automatic derivation of decadal historical patterns of population and dwellings at the level of a 100-m-mesh with commonly geodata from national mapping and cadastral agencies. High resolution population distribution maps were constructed by various dataset based on different methods (Dmowska & Stepinski, 2017; Gaughan et al., 2013; Linard, Gilbert & Tatem, 2011) like dasymetric redistribution (Rubinyi, Blankespoor & Hall, 2021; Sinha et al., 2019). And for the type of population, some models model residential population (Dobson et al., 2000; Fang & Jawitz, 2018) and some focus on ambient population (Bhaduri et al., 2007). However, none of them used the household as unit so the distribution of households was not considered. Meanwhile, Zhou et al. (2022) included both households and individuals as agents in their agent-based microsimulation. But they chose their typology of households (six types) best represented the population in their study area (Singapore).

Furthermore, UrbanSim is a popular tool that has been applied to several large cities in the United States for the microsimulation of urban land use and transportation, which provides household estimates as one of its functions (Waddell, 2002). However, the main purpose of this model is not to forecast population and household data; therefore, it does not consider the attributes of individual household members so cannot be used to aggregate the data at the household-level and obtain the population distribution according to age and gender. Moreover, UrbanSim has not yet been applied in Japan.

Therefore, the purpose of this research is to develop high-precision national-scale microsimulations of the future population and household distribution in Japan. Specifically, the household is set as the agent, and the address, household type, and five-year age groups and gender of each household member are estimated. The agent-based model was chosen due to the widespread availability of high-performance computing resources and large data storage capability (Zhou, Li, Basu & Ferreira, 2022) in this research. In addition, because the simulation requires current household data as the initial input, the creation of household estimation data for 2015 is set as a sub-target.

The remainder of the paper is organized as follows. Section 2 describes the household estimation data, which are prepared from national census and residential map, and the estimation method. Data obtained for Toyama and Shizuoka Prefectures from the national census are used to verify the accuracy of the model. Section 3 describes the construction of the household transition model, whose accuracy is verified by comparing the simulation results from 1980 to 2010 with corresponding data from the national census, as well as the results of simulations conducted from 2015 to 2045. In Section 4, the model is applied to predict the future distribution of disappearing villages and vacant houses by aggregating the simulation results at the village and building levels. Section 5 presents the conclusions of this study.

## 2. Preparation of estimated household data

Data on the number of household members, family type, relationship to householder, and age and gender of each household member were estimated from national census and vital statistics data. For the 2015 data, dwelling information of each household was also estimated from 2016 building data (Zmap TOWN, Zenrin Co., LTD.). The method of Akiyama et al. (2013) was improved and applied to the whole of Japan. The 1980 data were created only for Toyama and Shizuoka Prefectures in order to verify the accuracy of the household transition model.

### 2.1. Data

Table 1 summarizes the data used for the estimation of household, which were predominantly based on national census and vital statistics data published in e-Stat (National Statistics Center, 2021). However, data from the 1980 census (for each research region) were obtained from JoRAS, which is operated by the Center for Spatial Information Science at the University of Tokyo. Building data from 2016 were used in this study because the 2015 version had not yet been developed.

Table 1 Household estimation data used in this study

| Data | 2015 version | 1980 version |
|---|---|---|
| Building data | Zmap TOWNII, 2016 (Zenrin Co., Ltd.) | |
| Number of households by housing type | Housing and land survey of Japan, 2018 (municipality level, MIC) | |
| Number of vacant houses by housing type | National census, 2015 (municipality level, MIC) | National census, 1980 (municipality level, MIC) |
| Number of households by housing type & housing ownership | | |
| Number of households by housing ownership & family type | | |

| Number of households by family type & number of household members | | Create based on National census, 1980 (prefecture level, MIC) |
|---|---|---|
| Number of households by family type & age of householder & gender of householder | | |
| Age of husband & age of wife | National census, 2015 (Japan, MIC) | National census, 1980 (Japan, MIC) |
| Birth number by age of mother & birth order | Vital statistics, 2015 (Japan, MIC) | |
| Number of households by family type | National census, 2015 (subregion level, MIC) | National census, 1980 (research region level, MIC) |
| Number of households by number of household members | | |
| Male and female population in five-year age group | | |

## 2.2. Estimation method

This section describes the process for estimating household data. First, households were generated from the number of households in each municipality. Then, houses in the same town and village were designated using building data and attached to the location information. In the event of house shortages in some areas, other buildings were designated as houses. The attribute value was allocated to the generated household data according to the cross-tabulation data shown in Figure 1. For example, when the number of households owning a detached house was 30 and the number of households renting a house was 20, there was a 3/5 probability that a household living in a detached house owned the house. Similarly, the attributes of family type ($P(f|r)$), number of household members ($P(C|f)$), and age and gender of the householder ($P(A, G|f)$) were all allocated by probability values according to the cross-tabulation data. The number of households by family type and number of household members and the population by gender and age were adjusted for consistency for each town and village. For family types 5, 6, 7, 8, 11, 12, in which a couple and their parents live together (Table 2), there were no statistical data to distinguish whether the householder belonged to the couple generation or the parent generation. Because the accuracy of the data varied greatly depending on the generation of the householder, a sensitivity analysis was conducted. Finally, the households were sorted according to the age of the householder, under the assumption that the top 50% of households were headed by the parent generation. Table 2 lists the family types and their member compositions.

Next, the age and gender of the householder's spouse ($P(A'|A, G)$) were estimated from cross-tabulation of the couple's age using national data. For children, grandchildren, and parents in generations other than that of the married couple, the age of the householder was estimated using the number of births by the age of the mother and the birth order (vital statistics). Then, the age and gender of the householder, spouse, children, and grandchildren were estimated. Finally, the results of the previous estimates were compared with the population data by gender and age for each town and village, and the number of "other" household members was estimated to compensate for any gaps in the age and gender data.

In the 1980 version, the statistical data from the survey area was used instead of the data for each town and village, as shown in Figure 1. The cross-tabulation data of family type × household members and family type × age of householder × gender of householder were not published by municipality; therefore, data estimated for each prefecture were used instead. As building data could not be obtained for that year, the location of households could not be estimated accurately using only the municipality to which they belonged. In this study, the main improvements from the method of Akiyama et al. (2013) involved considering the generation of householder, using data for each town and street, and more effectively determining the attributes of "other" family members.

Table 2 Definition of family types and household members

| Family type I | | | | | |
|---|---|---|---|---|---|
| No. | Family type | Spouse | Children | Parent | Others |
| 1 | Couple | 1 | 0 | 0 | 0 |
| 2 | Couple + Children | 1 | 1–9 | 0 | 0 |
| 3 | Father + Children | 0 | 1–9 | 0 | 0 |
| 4 | Mother + Children | 0 | 1–9 | 0 | 0 |
| 5 | Couple + Parents | 1 | 0 | 2 | 0 |
| 6 | Couple + Parent | 1 | 0 | 1 | 0 |
| 7 | Couple + Children + Parents | 1 | 1–9 | 2 | 0 |
| 8 | Couple + Children + Parent | 1 | 9 | 1 | 0 |
| 9 | Couple + Others | 1 | 0 | 0 | 1–9 |
| 10 | Couple + Children + Others | 1 | 1–9 | 0 | 1 |
| 11 | Couple + Parents + Others | 1 | 0 | 1 | 1–9 |
| 12 | Couple + Children + Parents + Others | 1 | 1–9 | 1 | 1 |
| 13 | Brothers and sisters | 0 | 0 | 0 | 1–9 |
| 14 | Others | 0 | 0 | 0 | 1–9 |
| 15 | Nonrelative | 0 | 0 | 0 | 0 |
| 16 | Single | 0 | 0 | 0 | 0 |

| Family type II | | | | | |
|---|---|---|---|---|---|
| No. | Family type | Spouse | Children | Parent | Others |
| 5 | Couple + Parents | 1 | 0 | 2 | 0 |
| 6 | Couple + Parent | 1 | 0 | 1 | 0 |
| 7 | Couple + Children + Parents | 1 | 1–9 | 2 | 0 |
| 8 | Couple + Children + Parent | 1 | 9 | 1 | 0 |
| 11 | Couple + Parents + Others | 1 | 0 | 1 | 1–9 |
| 12 | Couple + Children + Parents + Others | 1 | 1–9 | 1 | 1 |

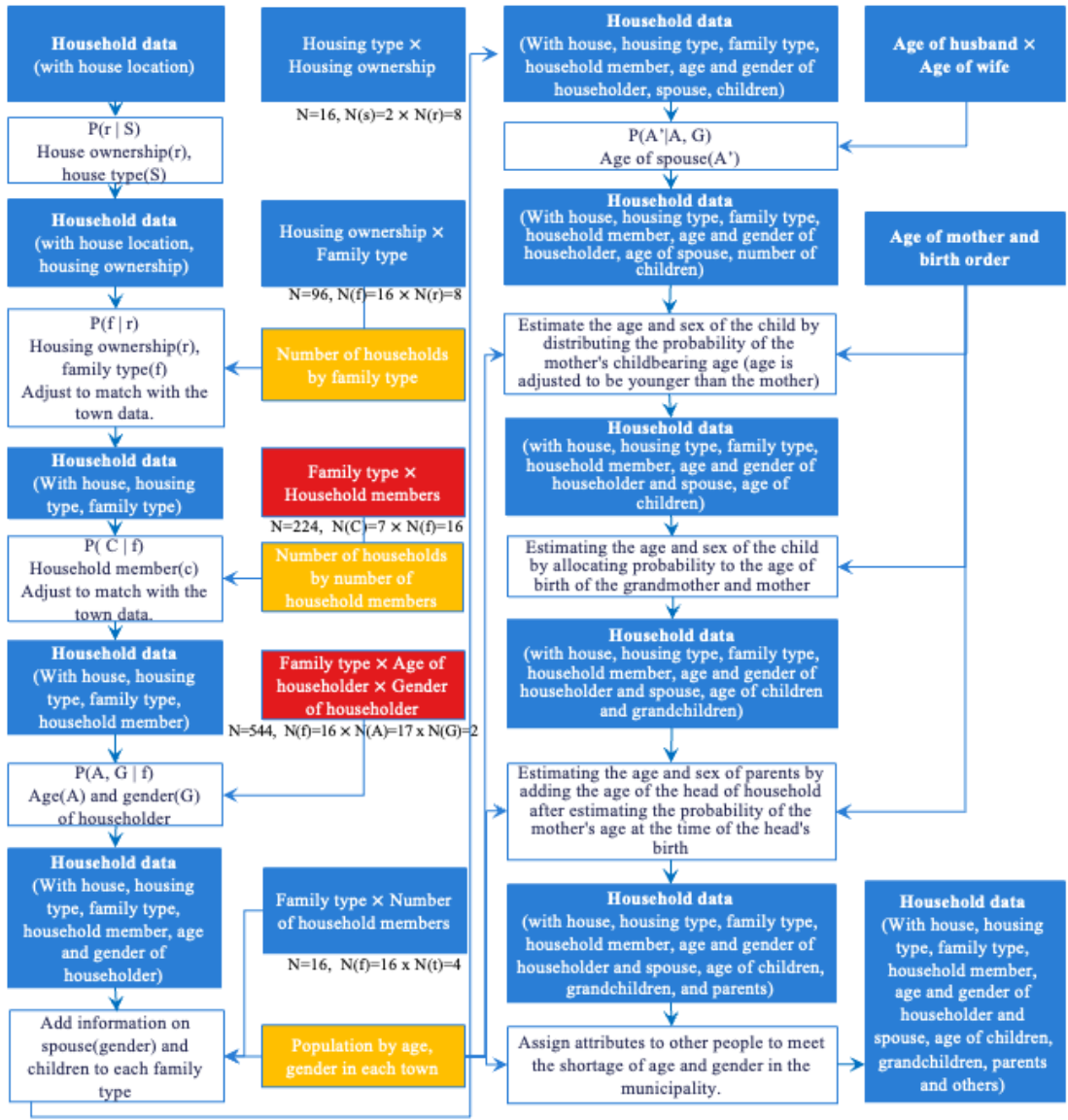

Fig. 1 Process of creating estimated household data

## 2.3. Verification

In this section, the accuracy of the 2015 household estimation data is verified. The accuracy of the 1980 data is omitted because the proposed method was optimized for the 2015 census data, whereas the 1980 data were created only to verify the accuracy of the household transition model. Furthermore, the age attribute data were edited to match that of the national survey. As shown in Table 3, the population by five-year age group, the number of households by family type (16 categories), and the population by town and village were compared with the national census data for each prefecture. The correlation coefficient (R) and mean absolute percentage error[3] (MAPE) were also calculated. The MAPE value was adopted because it is the most intuitive indicator for understanding the error between the actual measurement and the estimated value, where a lighter color in Table 3 indicates higher precision. Note that, for the estimated population data by town and village, the MAPE value was calculated by excluding towns and villages with fewer than 10 inhabitants because of their excessive influence on precision. The accuracy of the prefectural average is 0.921 (R value) & 6.135% (MAPE value) for population by five-year age group, 1.000 (R value) & 1.375% (MAPE value) for the number of households by family type, and 0.978 (R value) & 4.924% (MAPE value) for population by town and village. The error rate by housing type were −12.3% for detached houses and −1.4% for apartment houses. To verify the accuracy in more detail, the accuracy of the population by five-year age group and the number of households by family type (three categories) for each town and village were calculated using the average values for each prefecture. As shown in Table 4, the R value for the population aged 5–84 was generally 0.95 or higher and the MAPE value was less than 20%, whereas the same values for the number of households by family type (three categories) were 0.95 or higher and less than 12%, respectively, indicating high accuracy even in detailed areas. Conversely, for the population aged 0–4 and more than 85, the R and MAPE values were less than 0.8 and more than 30%, respectively, indicating low accuracy.

Furthermore, the estimation results reveal that the population aged 0–4 was particularly small in the Tokyo metropolitan area. This may be because of the method used to determine the generation of householders in three-generation families. In this study, 50% of households were assumed to be headed by the parent generation; however, this percentage may be higher in urban areas. Moreover, the generation of the householder varies greatly depending on the type of family, which was not surveyed in the national census. However, further improvements in accuracy can be expected if other statistical data are available. The reason for the significant error in the number of households by housing type is that the building data was not obtained from the same year as the census data (i.e., 2016 not 2015).

Otherwise, a comparison of the methods in the previous study by Akiyama et al. (2013) and this study with the 2010 household estimation data in Nanto, Toyama prefecture shows a significant improvement. Accuracy was ensured at the municipal level in terms of age distribution. And the accuracy was improved from the use of subregional data and in the distribution of family types by town and village.

Table 3 Accuracy of household estimation data (2015)

| Prefecture | Population of five-year age group | | Number of households by family type | | Population by town and village | | Error rate by housing type | |
|---|---|---|---|---|---|---|---|---|
| | R | MAPE (%) | R | MAPE (%) | R | MAPE (%) | Detached house | Apartment house |
| Average | 0.921 | 6.135 | 1.000 | 1.375 | 0.978 | 4.924 | −12.3 | −1.4 |
| Hokkaido | 0.914 | 7.101 | 1.000 | 2.271 | 0.982 | 5.328 | −15.8 | −1.4 |
| Aomori | 0.950 | 5.866 | 1.000 | 1.107 | 0.891 | 5.529 | −13.8 | −2.4 |
| Iwate | 0.917 | 6.479 | 1.000 | 0.999 | 0.953 | 4.440 | −12.8 | −2.1 |
| Miyagi | 0.921 | 7.034 | 1.000 | 1.293 | 0.929 | 4.157 | −20.9 | −1.9 |
| Akita | 0.935 | 6.997 | 1.000 | 0.989 | 0.987 | 5.577 | −6.8 | −1.3 |
| Yamagata | 0.898 | 5.990 | 1.000 | 0.885 | 0.991 | 4.988 | −8.8 | −0.6 |
| Fukushima | 0.915 | 7.072 | 1.000 | 0.572 | 0.978 | 4.296 | −11.9 | −1.7 |
| Ibaraki | 0.943 | 6.872 | 1.000 | 1.625 | 0.991 | 5.564 | −11.6 | −0.9 |
| Tochigi | 0.956 | 5.773 | 1.000 | 1.064 | 0.985 | 5.018 | −14.6 | −0.7 |
| Gunma | 0.946 | 5.286 | 1.000 | 1.550 | 0.985 | 4.496 | −12.2 | −0.3 |
| Saitama | 0.975 | 5.564 | 1.000 | 0.858 | 0.988 | 4.293 | −16.9 | −0.2 |
| Chiba | 0.961 | 6.629 | 1.000 | 0.683 | 0.989 | 4.231 | −15.0 | −0.6 |
| Tokyo | 0.984 | 4.815 | 1.000 | 0.583 | 0.982 | 2.713 | −14.3 | −0.1 |
| Kanagawa | 0.974 | 5.442 | 1.000 | 0.789 | 0.985 | 3.656 | −15.8 | −0.4 |
| Niigata | 0.895 | 6.509 | 1.000 | 1.642 | 0.988 | 3.956 | −9.6 | −0.8 |
| Toyama | 0.911 | 7.240 | 1.000 | 1.930 | 0.972 | 4.481 | −7.6 | −1.9 |
| Ishikawa | 0.895 | 7.187 | 1.000 | 2.362 | 0.981 | 5.988 | −8.9 | −0.8 |
| Fukui | 0.887 | 6.038 | 1.000 | 0.861 | 0.979 | 4.083 | −8.3 | −1.0 |
| Yamanashi | 0.925 | 5.406 | 1.000 | 1.910 | 0.980 | 5.443 | −14.8 | −0.3 |
| Nagano | 0.914 | 4.935 | 1.000 | 2.205 | 0.902 | 3.576 | −12.2 | −3.2 |
| Gifu | 0.924 | 6.109 | 1.000 | 1.987 | 0.931 | 3.751 | −21.3 | −5.5 |
| Shizuoka | 0.953 | 5.076 | 1.000 | 1.861 | 0.989 | 3.782 | −12.1 | −0.5 |
| Aichi | 0.964 | 5.786 | 1.000 | 1.633 | 0.993 | 3.063 | −15.4 | −0.5 |
| Mie | 0.913 | 6.628 | 1.000 | 1.489 | 0.980 | 4.915 | −18.1 | −1.3 |
| Shiga | 0.920 | 7.120 | 1.000 | 1.240 | 0.979 | 4.257 | −16.1 | −1.2 |
| Kyoto | 0.920 | 6.800 | 1.000 | 1.544 | 0.995 | 3.380 | −12.2 | −1.2 |
| Osaka | 0.969 | 5.317 | 1.000 | 0.805 | 0.987 | 3.510 | −12.7 | −0.2 |
| Hyogo | 0.940 | 6.226 | 1.000 | 0.808 | 0.995 | 3.395 | −12.0 | −0.6 |
| Nara | 0.933 | 6.535 | 1.000 | 1.093 | 0.978 | 6.044 | −10.6 | −1.7 |
| Wakayama | 0.936 | 5.007 | 1.000 | 2.426 | 0.989 | 5.875 | −17.4 | −3.2 |
| Tottori | 0.841 | 6.691 | 1.000 | 0.705 | 0.995 | 5.529 | −4.3 | −2.8 |
| Shimane | 0.872 | 6.753 | 1.000 | 1.839 | 0.987 | 6.954 | −8.1 | −1.4 |
| Okayama | 0.881 | 6.231 | 1.000 | 1.297 | 0.993 | 4.510 | −13.0 | −1.1 |
| Hiroshima | 0.908 | 6.393 | 1.000 | 1.277 | 0.990 | 4.017 | −18.3 | −0.5 |
| Yamaguchi | 0.907 | 6.723 | 1.000 | 1.842 | 0.990 | 4.895 | −14.5 | −0.9 |
| Tokushima | 0.920 | 5.890 | 1.000 | 2.816 | 0.997 | 4.789 | −9.4 | −1.0 |
| Kagawa | 0.936 | 5.316 | 1.000 | 2.805 | 0.975 | 8.167 | −9.1 | −2.0 |
| Ehime | 0.909 | 6.258 | 1.000 | 1.406 | 0.973 | 4.593 | −14.4 | −1.3 |
| Kochi | 0.904 | 6.129 | 1.000 | 1.663 | 0.986 | 8.786 | −8.0 | −1.9 |
| Fukuoka | 0.922 | 6.183 | 1.000 | 0.690 | 0.971 | 4.950 | −17.1 | −0.9 |
| Saga | 0.892 | 5.217 | 1.000 | 1.226 | 0.984 | 3.219 | −9.1 | −1.8 |
| Nagasaki | 0.895 | 6.366 | 1.000 | 1.054 | 0.979 | 4.880 | −6.8 | −1.8 |
| Kumamoto | 0.884 | 5.073 | 1.000 | 0.670 | 0.975 | 6.150 | −9.0 | −1.2 |
| Oita | 0.864 | 6.848 | 1.000 | 0.944 | 0.984 | 5.749 | −8.6 | −1.0 |
| Miyazaki | 0.885 | 6.583 | 1.000 | 1.070 | 0.994 | 4.287 | −6.6 | −1.1 |
| Kagoshima | 0.890 | 6.328 | 1.000 | 1.182 | 0.968 | 9.787 | −12.4 | −5.1 |
| Okinawa | 0.974 | 4.522 | 1.000 | 1.092 | 0.987 | 6.366 | −8.0 | −0.3 |

Note: lighter color indicates higher precision.

Table 4 Accuracy of estimated household data by family type and age (prefecture-level)

| Family type | Household | |
|---|---|---|
| | R | MAPE (%) |
| Core types | 0.982 | 2.3916 |
| Single-member | 0.988 | 2.5254 |
| Others | 0.956 | 11.512 |

| Age | Population | | Age | Population | |
|---|---|---|---|---|---|
| | R | MAPE (%) | | R | MAPE (%) |
| 0–4 | 0.572 | 65.147 | 45–49 | 0.977 | 10.385 |
| 5–9 | 0.957 | 19.064 | 50–54 | 0.975 | 10.792 |
| 10–14 | 0.969 | 13.834 | 55–59 | 0.971 | 12.171 |
| 15–19 | 0.938 | 11.975 | 60–64 | 0.967 | 14.314 |
| 20–24 | 0.954 | 10.137 | 65–69 | 0.964 | 15.862 |
| 25–29 | 0.971 | 10.445 | 70–74 | 0.958 | 17.335 |
| 30–34 | 0.976 | 11.693 | 75–79 | 0.95 | 18.257 |
| 35–39 | 0.978 | 11.315 | 80–84 | 0.937 | 19.357 |
| 40–44 | 0.978 | 10.493 | 85– | 0.762 | 34.88 |

## 3. Construction of the household transition model

A model was constructed to estimate the future population distribution of households, using the household as the agent, based on household estimation data for each municipality. In the estimation, three population change factors were considered: births, deaths, and migration. These factors are also considered in the cohort factor method, which is a typical method for estimating the future population, whereby all population changes are mediated by these three factors[2]. Furthermore, migration was subdivided into out-migration, in-migration, marriage, and divorce. Using the probability data of each factor, the change in households in each municipality was simulated every five years from 1980 to 2010 and from 2015 to 2045. Simulations for 1980–2010 were conducted to verify the accuracy of the model. Whereas the accuracy of the household estimation data was evaluated at the prefecture level for the whole of Japan, the accuracy of the household transition model was evaluated at the municipality level in Toyama and Shizuoka Prefectures for more local accuracy. As building data were not available for 1980, the accuracy of household location was not verified.

## 3.1. Data

As shown in Table 5, Zmap TOWN II (Zenrin Co., Ltd.) data for building data, future assumed values determined by the National Institute of Population and Social Security Research (IPSS) for survival rate and in/out-migration rate data, and national census data were utilized for the 2015–2045 simulation. Statistical data from the 2015 census were also used. For the 1980–2010 simulation, statistical census data from 2015 were used exclusively, and annual probabilities were adopted. The future assumed values from IPSS are obtained by estimating future population by region and are available on their website as of January 2021 (National Institute of Population and Social Security Research, n.d.). Values were calculated according to population changes from 2010 to 2015, and only for those areas that deviated greatly from the population before 2010, by comprehensively considering past trends. For migration rate data, only the migration rate (subtracting the out-migration rate from the in-migration rate) needs to be calculated. Thus, the out-migration rate was determined using the value calculated from 2010–2015 census data, and the in-migration rate was obtained by adding the migration rate to the out-migration rate for each five-year period during 2015–2045. In addition, census data related to population movement were released only every 10 years prior to 2010. Therefore, the migration rate statistics for 1990, 2000, and 2010 were used to calculate the migration rate in 1985, 1995, and 2005, respectively.

Table 5 Data used to construct the household transition model

|  | 2015–2045 | 1980–2010 |
|---|---|---|
| Building data | Zmap TOWN II 2016 (Zenrin Co., Ltd.) |  |
| Survival rate (by gender, five-year age group, and city) | Future assumed value (IPSS) | National census 1985–2010 (MIC) |
| In/out-migration rate (by gender, five-year age group, and city) | National census 2015 (MIC) Future assumed value (IPSS) | National census 1980, 1990, 2000, 2010 (MIC) |
| In/out-migration rate within prefecture (by gender, five-year age group, and city) | National census 2015 (MIC) |  |
| Distribution of out-migration households (by prefecture) | National census 2015 (MIC) |  |
| Households in start year of simulation (by family types, city) | National census 2015 (MIC) |  |
| In/out migration households by family types (by prefecture) | National census 2015 (MIC) |  |
| In/out migration by small region | National census 2015 (MIC) |  |
| Marriage rate (female, by five-year age group, and prefecture) | National census 2015 (MIC) | National census 1985–2010 (MIC) |
| Distribution of couples' age (by gender, five-year age group, nationwide) | National census 2015 (MIC) | |
| Rate of living together with parents (by prefecture) | National census 2015 (MIC) | |
| Divorce rate (female, by five-year age group, and prefecture) | National census 2015 (MIC) | |
| Rate of divorce with/without child (nationwide) | National census 2015 (MIC) | |
| Custody rate after divorce (nationwide) | National census 2015 (MIC) | |
| Natality (by city) | Future assumed value (IPSS) | |
| Natality by birth order (female, by five-year age group, nationwide) | National census 2015 (MIC) | |
| Proportion of births to legitimate and illegitimate children (nationwide) | National census 2015 (MIC) | |
| Gender ratio for age between 0–4 (by city) | Future assumed value (IPSS) | |

## 3.2. Simulation procedure

Figure 2 shows the simulation procedure of 2015-2045. Changes in households through deaths, move-out, move-in, marriages, divorces, and births were simulated every five years. Household estimation data were simulated for each municipality using the household as the agent and considering movement between municipalities within a prefecture with strong interrelationships. The address of households was not considered in the 1980–2010 simulation due to the lack of building data but was considered in the 2015–2045 simulation. To verify the accuracy of the model, the age and gender in the household estimation data were edited to fit the actual measurement data for data of 1980, which was employed as the base year. Each step of the simulation is explained in detail below.

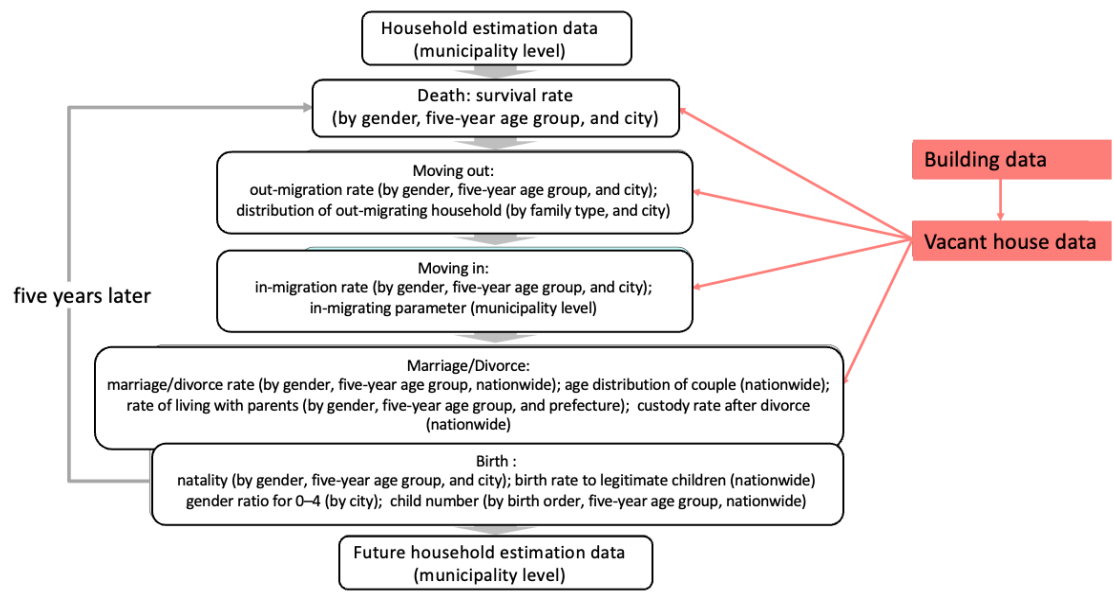

Fig. 2 Simulation procedure of the household transition model (2015-2045)

1) Creation of vacant house data

Prior to the start of the simulation process, a list of vacant houses in 2015 was created by subtracting the dwellings used in the household estimation data from the dwellings in the building data. Furthermore, to bring the number of vacant houses in the entire municipality closer to the actual value, the number of households with dwellings being designated as offices and targets in the household estimation data were randomly subtracted from the remaining buildings. A household was deleted or added every time a household was created or destroyed in an event.

2) Deaths

Deaths of household members for each household agent were added according to the survival rate (by gender, five-year age group, and city). Households where all members had died were added to the list of vacant houses.

3) Moving out

The population distribution of out-migrants for each city and out-migrants within the prefecture (by gender, five-year age group, and city) were determined from the out-migration rate and out-migrants rate within the prefecture (by gender, five-year age group, and municipality). Then, households that had moved out were determined according to the distribution of out-migrating households by family type (prefecture base), the distribution of migrating households by housing type (prefecture base), and the number of in-migrants by small region. The counts were subtracted from the distribution of the out-migrating population and the out-migrating population within the prefecture (by gender, five-year age group, municipality). The number of in-migrants for each small region was employed because the number of out-migrants in small regions was not available as statistical data, and the trends of moving in and out are assumed to be consistent. As the entire population of the prefecture (by gender, five-year age group, municipality) cannot be transferred out by households, the remaining members were moved out individually from each household. Then, the households that had moved out from each municipality within the prefecture were integrated into a dataset of individual moving data within the prefecture and stored for the next step. Finally, the households and individuals that had moved out were determined in the same way until there were no more individuals left in the out-migrant distribution data (by gender, five-year age group, municipality). In this scenario, out-migrant data were not stored. New vacant houses were added to the dataset.

4) Moving in

The population distribution of in-migrants and number of in-migrants within the prefecture (by gender, five-year age group, municipality) were determined using the same method as the moving-out scenario, based on the in-migration rate and number of in-migrants within the prefecture (by gender, five-year age group, municipality). Households were selected from the in-prefecture moving household data saved in the previous step according to the distribution of the population moving into the prefecture (by gender, five-year age group, municipality), and added as moving-in households. The remaining households were classed as moving-in households according to the distribution of in-migrants (by gender, five-year age group, municipality). These moving-in households were equipped with the location selected from vacant house data according to the distribution of moving-in households by housing type (by prefecture) and the number of in-migrants by small region. The method of assigning housing addresses was the same as that for new households created by marriage or divorce.

5) Marriage

Because moving in and out as a result of marriage are included in the moving-in and moving-out rates, only marriages within the city are considered in this scenario. The marriage rate (female, by five-year age group and prefecture) was used to select women who marry from among unmarried women. Unmarried men were selected from the city based on the age distribution of married couples (by gender, five-year age group, nationwide). Couples were then determined using the selected men and women. Finally, whether the couple then live with their parents or together was decided according to the rate of cohabitation with parents (by prefecture). For couples living with parents, the probability of living with either set of parents was considered equal. Those couples who started to live together were located in vacant houses.

6) Divorce

The divorce simulation was performed according to data on women, that is, divorcing couples were determined according to the divorce rate (female, by five-year age group and prefecture) and the divorce rate with and without children (nationwide). For couples with children, the children were randomly assigned to one side of the family at the same rate. The remaining single person was located in a vacant house.

7) Birth

Women between 15 and 50 years old were assumed to be able to give birth to children. Those predicted to have children were determined according to several multiple conditions including natality (by municipality), natality by birth order (female, by five-year age group, nationwide), the proportion of births of legitimate to illegitimate children (nationwide), marital status, and whether or not the woman has already had children. The gender of the child was determined according to the gender ratio of children aged 0–4 (by municipality).

The simulation was completed by repeating steps (2) to (7) six times over 30 years (five-year period for once).

**3.3. Verification**

Simulations were conducted for all municipalities in Toyama and Shizuoka Prefectures for 1980–2010, then compared with the population by five-year age group and the number of households by family type (16 categories) from the 2010 national census data (National Statistics Center, 2021). As shown in Table 6, the average accuracy of the population by five-year age group and municipality was high (R = 0.993; MAPE = 1.584%). For the number of households by family type (16 categories), R was 0.975 and MAPE was 19.39%. As the error in the number of households by family type is not negligible, family types that may affect the accuracy were further examined. The 16 family types were divided into three categories: core-type families (family types 1–4), non-core-type families with only relatives (family types 5–15), and single-member families (family type 16). Figure 3 shows the errors in the simulation results based on the 2010 census data and counted by municipality.

The accuracy of the estimates tended to decrease from core-type families to non-core-type families with only relatives, with single-member families exhibiting the lowest accuracy. In many municipalities, the accuracy was overestimated by 0–30% for core-type families, whereas the errors for non-core-type families ranged from −50% to 10%. For single-member families, the error ranged between −30% and 40%. The underestimation of non-core-type families with only relatives is due to the fact that the behavior of three-generation families could not be accurately simulated. Three-generation families, which account for a large proportion of non-core-type families with only relatives, are formed when couples live with their parents after marriage and have children. The rate of living together with parents used in this study was based on statistical data for each prefecture; however, there may be insignificant regional differences even within a prefecture. Furthermore, the error for single-member families is related to the unstable status of this family type. Many life events such as marriage and moving in and out effect this family type, which reduces the accuracy of simulations.

Table 6 Simulation accuracy by municipality in 2010 (Toyama/Shizuoka Prefecture)

| Municipality | Population by five-year age group | | Number of households by family type | | Municipality | Population by five-year age group | | Number of households by family type | | Municipality | Population by five-year age group | | Number of households by family type | |
|---|---|---|---|---|---|---|---|---|---|---|---|---|---|---|
| | R | MAPE (%) | R | MAPE (%) | | R | MAPE (%) | R | MAPE (%) | | R | MAPE (%) | R | MAPE (%) |
| **Average** | **0.993** | 1.5842 | **0.975** | 19.393 | Shizuoka | **0.986** | 2.2208 | 0.994 | 12.782 | Fukuroi | 0.998 | 1.1656 | 0.973 | 20.989 |
| Toyama | 0.994 | 1.5571 | 0.976 | 13.291 | Hamamatsu | 0.984 | 1.7629 | 0.994 | 15.454 | Izunokuni | 0.999 | 1.186 | 0.988 | 11.505 |
| Takaoka | 0.998 | 1.2765 | 0.976 | 15.552 | Numazu | 0.999 | 1.102 | 0.992 | 14.957 | Susono | 0.969 | 2.9235 | 0.965 | 31.615 |
| Uozu | 0.998 | 1.4194 | 0.983 | 14.371 | Atami | 0.998 | 2.194 | 0.969 | 21.333 | Kosai | 0.981 | 3.0641 | 0.988 | 22.415 |
| Himi | 0.999 | 1.0394 | 0.979 | 18.87 | Mishima | 0.997 | 1.2367 | 0.988 | 26.213 | Izu | 0.998 | 1.2795 | 0.967 | 14.881 |
| Kurobe | 0.999 | 1.0206 | 0.986 | 11.063 | Fujinomiya | 0.997 | 1.4219 | 0.974 | 21.315 | Omaezaki | 0.987 | 2.1675 | 0.981 | 17.912 |
| Namerikawa | 0.996 | 1.845 | 0.984 | 13.556 | Ito | 0.995 | 2.3925 | 0.978 | 17.382 | Kikugawa | 0.998 | 0.9536 | 0.968 | 20.703 |
| Tonami | 0.999 | 0.6791 | 0.969 | 16.608 | Shimada | 0.999 | 0.9608 | 0.961 | 21.708 | Izunokuni | 0.994 | 2.3122 | 0.963 | 26.87 |
| Oyabe | 0.991 | 2.6338 | 0.951 | 19.385 | Fuji | 0.999 | 1.1687 | 0.975 | 21.095 | Makinohara | 0.998 | 0.9717 | 0.967 | 20.674 |
| Nanto | 0.998 | 1.4373 | 0.969 | 14.104 | Iwata | 0.977 | 2.01 | 0.981 | 20.948 | Kamo | 1.000 | 0.9779 | 0.997 | 9.7164 |
| Imizu | 0.999 | 0.8558 | 0.967 | 17.148 | Yaizu | 0.999 | 1.1389 | 0.962 | 23.689 | Tagata | 0.999 | 1.0053 | 0.929 | 37.439 |
| Nakaniikawa | 0.998 | 1.3578 | 0.994 | 7.997 | Kakegawa | 0.988 | 1.4903 | 0.983 | 17.038 | Sunto | 0.991 | 2.5037 | 0.970 | 29.767 |
| Shimoniikawa | 0.990 | 2.49 | 0.961 | 24.419 | Fujieda | 0.997 | 1.2636 | 0.967 | 23.355 | Haibara | 0.950 | 2.0659 | 0.979 | 20.005 |
| | | | | | Gotemba | 0.992 | 1.6378 | 0.961 | 32.499 | Shuchi | 0.998 | 1.1798 | 0.972 | 15.094 |

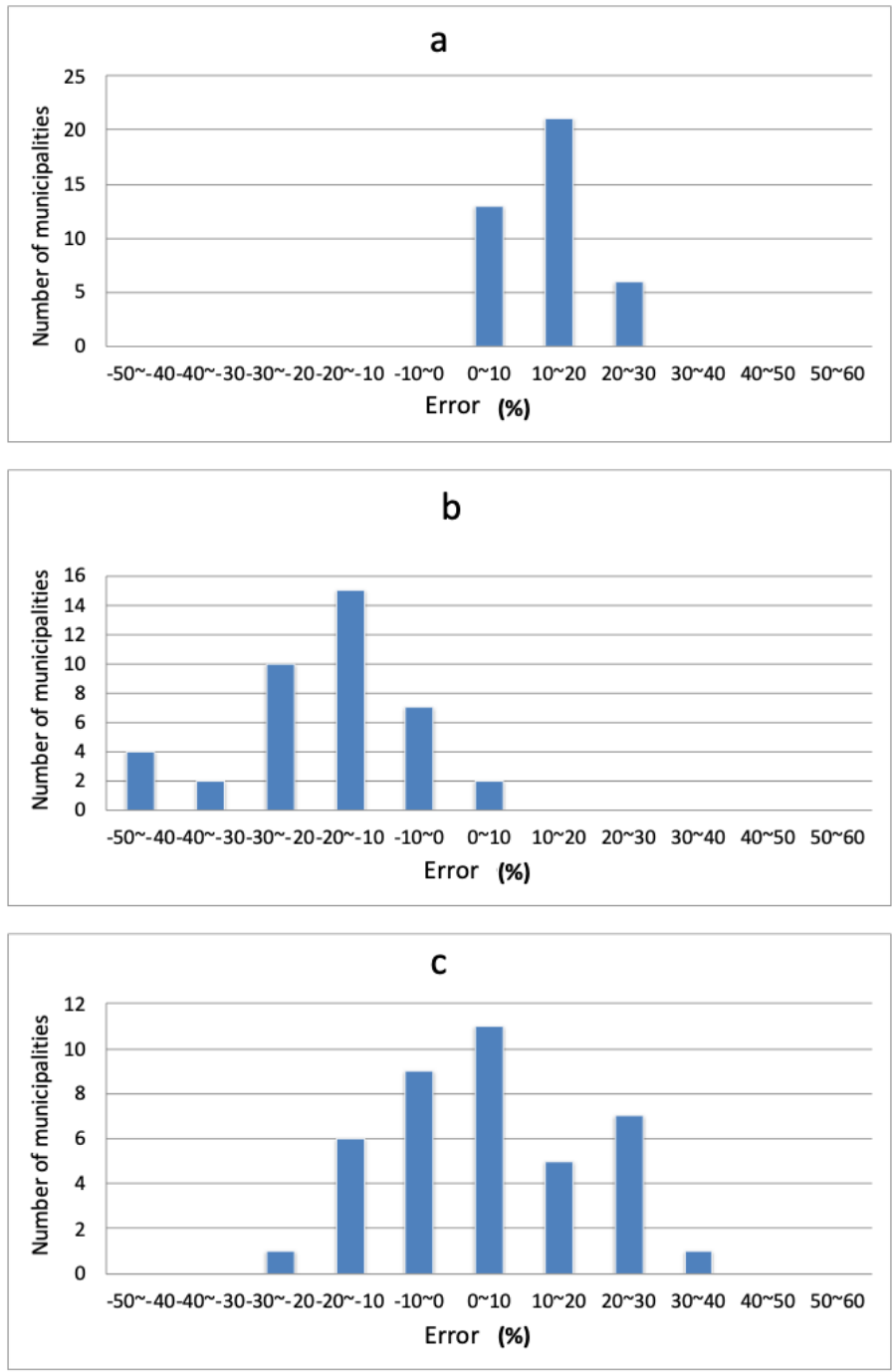

Fig. 3 Error distribution of the number of households by family type: (a) core-type families (family types 1–4), (b) non-core-type families with only relatives (family types 5–15), (c) single-member families (family type 16)

## 4. Applications of the household transition model

Here, the household transition model was applied to forecast the distribution of vacant houses and disappearing villages, which are typical regional issues for local governments in Japan.

## 4.1. Forecasting vacant houses

The distribution of vacant houses was estimated in Nanto City, Toyama Prefecture using household and building data with the household transition model. The number and distribution of vacant detached houses in 2020–2040 were calculated based on household estimation data for 2010 and compared with actual data, as shown in Table 7. A survey of vacant houses was conducted by Nanto City Hall staff through interviews and analyses of building appearance. According to the household transition model, the number of vacant houses in 2020 was 1,197, which is close to the values determined by the 2018 housing and land survey (1,450) and the 2017 vacant house survey (1,035). The number of vacant houses is predicted to rapidly increase in 2040 because the baby boomer generation will be in their 90s at this time, and will therefore exhibit a high mortality rate. Furthermore, detailed data from the 2017 Nanto City vacant house survey were aggregated into a 500-m mesh to reveal the geographical distribution of vacant detached houses, which is compared to the results of the 2020 household transition model in Figure 4. The results indicate that the proposed model accurately reproduces the areas where vacant houses are concentrated in Nanto City.

Table 7 Comparison of simulated and actual vacant house (detached house) data in Nanto City, Toyama Prefecture

| Year | Household transition model | | Housing/Land survey | | Vacant house survey |
|---|---|---|---|---|---|
| | Household | Vacant house (detached house) | Household | Vacant house (detached house) | |
| 2010 | 16887 | 881 | | | |
| 2013 | | | 18450 | 1520 | |
| 2015 | 16887 | 1178 | | | |
| 2016 | | | | | 902 |
| 2017 | | | | | 1035 |
| 2018 | | | 18380 | 1450 | |
| 2020 | 16887 | 1197 | | | |

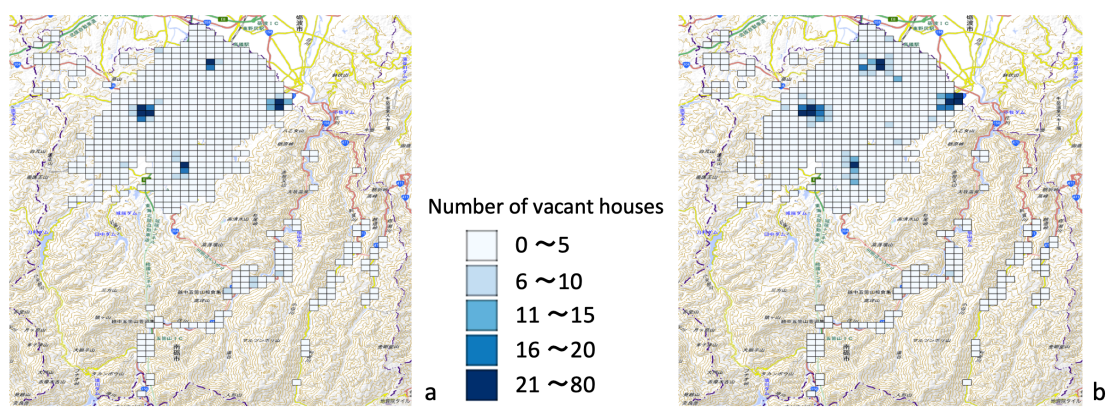

Fig. 4 Comparison of the distribution of vacant detached houses (500-m mesh) determined by (a) the 2017 Nanto City vacant house survey and (b) the 2020 household transition model

## 4.2. Forecasting disappearing villages

The household transition model was used to estimate the number of villages that will disappear in the Hokuriku region of Japan based on the number of households. Boundary data of agricultural communities used in the model were collected from the Ministry of Agriculture, Forestry, and Fisheries' Census (2015). Agricultural villages are defined as a basic unit of social life in a naturally occurring village, where households are connected to each other by geographical and blood ties, and where various groups and social relationships have been formed (Ministry of Agriculture, Forestry and Fisheries, 2015). These data were considered appropriate for this study because they cover not only agriculture but also local social villages. Although there is no clear standard for the number of households and population that can sustain a village, it can be assumed that a village will disappear when there are fewer than five households and less than 10 people. Agricultural villages in the Hokuriku region were simulated from 2015 to 2045 and the results were compared with data from the "Report on the Current Status of Villages in Depopulated Areas" of the Ministry of Internal Affairs and Communications (MIC, 2020). The average number of extinct villages between determined by MIC was 2.17 for 2015–2019, whereas that determined by the proposed model was 3.4 for 2015–2020. Figure 5 shows the average number of extinct villages between 2015 and 2045, as predicted by the proposed model, which indicates that villages are predominantly disappearing in mountainous areas and the Noto Peninsula.

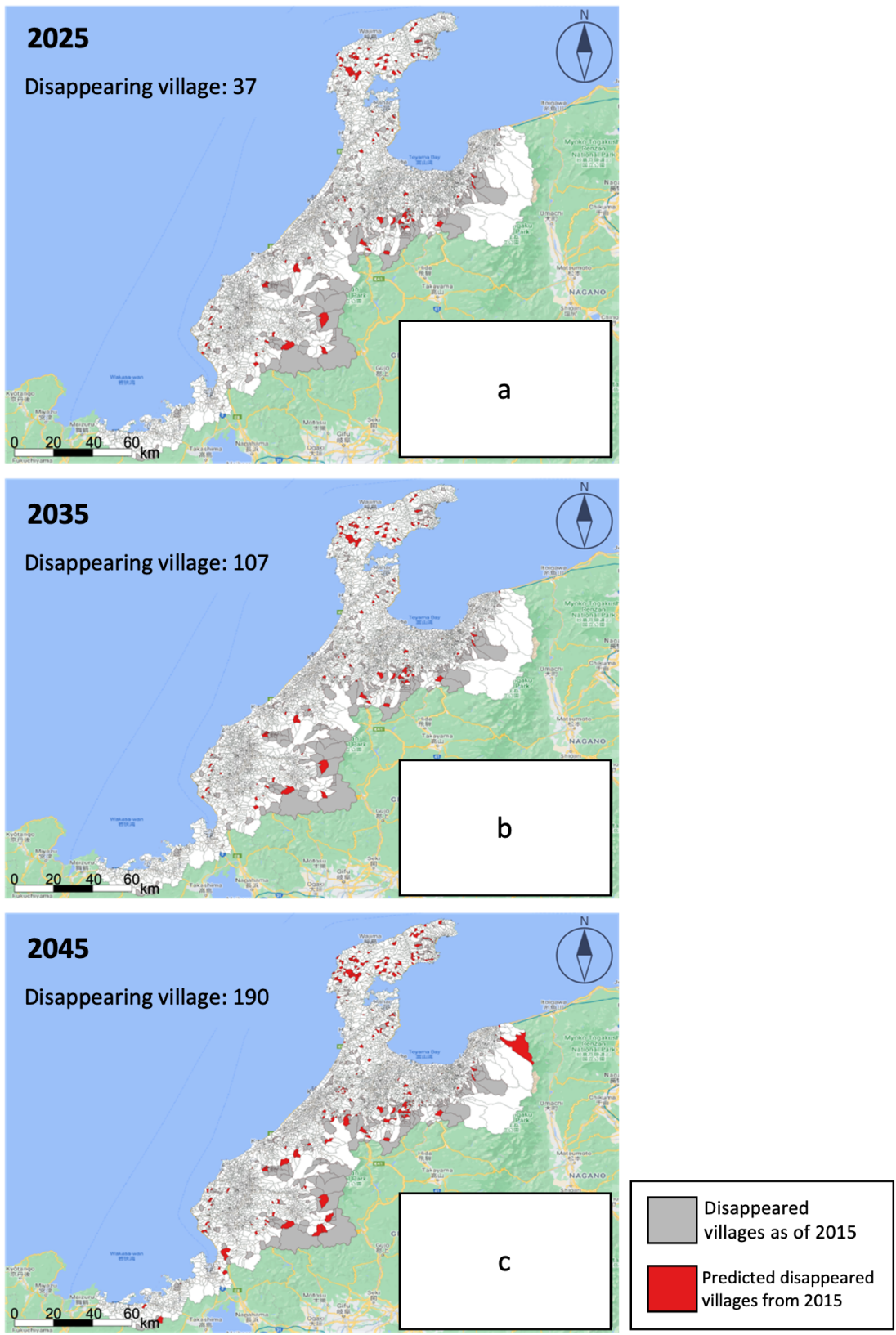

Fig. 5 Simulation results of disappearing villages in the Hokuriku Region of Japan: (a) 2025, (b) 2035, (c) 2045

## 5. Conclusions

In this study, a household transition model was constructed using household estimation data for the whole of Japan in 2015, then used to forecast the distribution of vacant houses and disappearing villages in parts of Japan. The household estimation data exhibited high accuracy for both the population by subregion, gender, and five-year age group and the number of households by municipality and 16 family types. Therefore, the results of this study can be used not only in long-term simulations, but also in a wide range of fields, such as research on human flows to estimate daily activities. These data can also be updated as new census data are released. The accuracy of the household transition model was verified by simulating the population and number of households in Toyama and Shizuoka Prefectures for 1980–2010. High accuracy was confirmed for both the population by city, gender, and five-year age group and the number of households by city and 16 family categories. No previous studies have conducted simulations that focus on the age and gender of individual household members; thus, this study is the first to establish a microsimulation of households. Moreover, the proposed model estimated future distributions of vacant houses and disappearing villages that were consistent with data from various surveys.

Nevertheless, this study has some limitations. In the household estimation data, the number of dwellings does not correspond to that of the national census because of the lack of building data for 2015. Moreover, determining the generation of the householder in three-generation families contains considerable uncertainty. Thus, the accuracy of the model can be improved in future by solving these problems. Unfortunately, the accuracy of changes in the detailed geographical distribution of households over time cannot be successfully verified by the household transition model. That is, although household estimation data can be generated for 2015 and the corresponding building data are currently available, the accuracy of the medium-term household transition over a period of 15 years can only be verified when the results of the 2020 census are released. In addition, the accuracy of in/out-migration also needs to be improved. Detailed data of in/out-migration by households are not available, whereas the statistical data was available so in/out-migration parameters were applied in current model. The family types of in/out-migration were also limited to 1, 2, 4, 6, 7 and 16, which was not always realistic. In summary, although the simulation results are consistent with the data of various surveys, the accuracy of the proposed household transition model can be further improved in future.

**References**

Abe, Y. (2011). Family labor supply, commuting time, and residential decisions: The case of the Tokyo Metropolitan Area. *Journal of Housing Economics*, 20(1), 49–63.

Akiyama, Y., Takada, H., & Shibasaki, R. (2013). Development of micropopulation census through disaggregation of national population census. *CUPUM 2013 Conference Papers*, 110, 1–33. Retrieved from https://cupum2013.geo.uu.nl/download/usb/contents/pdf/shortpapers/110_Akiyama.pdf. Accessed March 1, 2021.

Akkerman, A. & Shimoura, S. (2012). Discrete choice in commuter space: Small area analysis of diurnal population change in the Tokyo Metropolitan Region. *Computers, Environment and Urban Systems*, 36, 386–397.

Ballas, D., Clarke, G., Dorling, D., Eyre, H., Thomas, B. & Rossiter, D. (2005), *SimBritain*: a spatial microsimulation approach to population dynamics. *Population, Space and Place*, 11, 13–34.

Bhaduri, B., Bright, E., Coleman, P. & Urban, ML. (2007). LandScan USA: A high-resolution geospatial and temporal modeling approach for population distribution and dynamics. *GeoJournal*, 69, 103–117.

Center for Spatial Information Science, The University of Tokyo. (n.d.) *JoRAS*. Joras.csis.u-tokyo.ac.jp. Retrieved from https://joras.csis.u-tokyo.ac.jp. Accessed March 1, 2021.

Chikuma, M. & Sato, T. (2017). Development of the population distribution estimation method in a city to examine the measures to attract residence based on the location optimization plan and a case study for Toyohashi city, Japan. *Journal of the City Planning Institute of Japan*, 52(3), 1124–1129.

Christiansen, S. & Keilman, N. (2013). Probabilistic household forecasts based on register data–the case of Denmark and Finland. *Demographic Research*, 28, 1263–1302.

Dmowska, A. & Stepinski, T. (2017). A high resolution population grid for the conterminous United States: The 2010 edition. *Computers, Environment and Urban Systems*, 61, 13-23.

Dobson, JE., Bright, EA., Coleman, PR., Durfee, RC. & Worley, BA. (2000). LandScan: a global population database for estimating population at risk. *Photogramm Eng Rem S*, 66, 849–857.

Fang, Y. & Jawitz, J. (2018). High-resolution reconstruction of the United States human population distribution, 1790 to 2010. *Sci. Data*, 5: 180067.

Gaughan, A. E., Stevens, F. R., Linard, C., Jia, P. & Tatem, A. J. (2013). High resolution population distribution maps for Southeast Asia in 2010 and 2015. *PLoS ONE*, 8, e55882.

Hasegawa, Y., Sekimoto, Y., Seto, T., Fukushima, Y., & Maeda, M. (2019). My City Forecast: Urban planning communication tool for citizen with national open data. *Computers, Environment and Urban Systems*, 77, 101255.

Hattori, K., Kaido, K. & Matsuyuki, M. (2017). The development of urban shrinkage discourse and policy response in Japan. *Cities*, 69, 124–132.

Hecht, R., Herold, H., Behnisch, M. & Jehling, M. (2019). Mapping Long-Term Dynamics of Population and Dwellings Based on a Multi-Temporal Analysis of Urban Morphologies. *ISPRS Int. J. Geo Inf., 8*, 2.

Karashima, K., Ohgai, A. & Saito, Y. (2014). A GIS-based support tool for exploring land use policy considering future depopulation and urban vulnerability to natural disasters – a case study of Toyohashi City, Japan. *Procedia Environmental Sciences*, 22, 148–155.

Keilman, N. (2016). Household forecasting: Preservation of age patterns. *International Journal of Forecasting*, 32(3) 726–735.

Keito, I. & Hao, L. (2019). Population estimation microsimulation with location selection model–a case study of Hamamatsu City. *Journal of the Japan Society for Management Information*, 201910, 58–61.

Lee, J., Akashi, Y., Takaguchi, H., Sumiyoshi, D., Lim, J., Ueno, T., Maruyama, K. & Baba, Y. (2021). Forecasting model of activities of the city-level for management of $CO_2$ emissions applicable to various cities. *Journal of Environmental Management*, 286, 112210.

Linard, C., Gilbert, M. & Tatem, A. J. (2011). Assessing the use of global land cover data for guiding large area population distribution modelling. *GeoJournal*, 76, 525–538.

Masuda, H. (2014), Chiho Shometsu: Tokyo ikyoku shuchu ga maneku jinko kyugen [Extinction of Rural Municipalities: Rapid population decline in rural Japan caused by unipolar centralization in Tokyo] (in Japanese), Chuokoron-Shinsha, Inc.

Matsunaka, R., Oba, T., Nakagawa, D., Nagao, M. & Nawrocki, J. (2013). International comparison of the relationship between urban structure and the service level of urban public transportation–A comprehensive analysis in local cities in Japan, France and Germany. Transport Policy, 30, 26–39.

Ministry of Agriculture, Forestry and Fisheries. (2015). *Explanation of terms used in the Census of Agriculture and Forestry, etc.* Retrieved from https://www.maff.go.jp/j/study/census/2015/1/pdf/sankou5.pdf. Accessed March 1, 2021.

Ministry of Internal Affairs and Communications. (2020). *Survey report on the current status of settlements in depopulated areas, etc.* Retrieved from https://www.soumu.go.jp//000678497.pdf. Accessed March 1, 2021.

Ministry of Land, Infrastructure, Transport and Tourism. (2021). *Status of preparation of location normalization plan*. Retrieved from https://www.mlit.go.jp/toshi/city_plan/content/001385755.pdf. Accessed March 1, 2021.

Miyauchi, T., Setoguchi, T. & Ito, T. (2021). Quantitative estimation method for urban areas to develop compact cities in view of unprecedented population decline. *Cities*, 114, 103151.

Münnich, R., Schnell, R., Brenzel, H., Dieckmann, H., Dräger, S., Emmenegger, J., Höcker, P., Kopp, J., Merkle, H., Neufang, K., Obersneider, M., Reinhold, J., Schalle, J., Schmaus, S. & Stein, P. (2021), A population based regional dynamic microsimulation of Germany: The MikroSim Model. *Methods, data, analyses*, 15(2), 241–264.

Murphy, M. (1991). Household modelling and forecasting–dynamic approaches with use of linked census data. *Environment and Planning A*, 23, 885–902.

National Institute of Population and Social Security Research. (2018). *Future population projections by region for Japan - 2015 to 2045 - (estimated in 2008)*. Tokyo, Japan: National Institute of Population and Social Security Research, 2–26. Retrieved from https://www.ipss.go.jp/pp-shicyoson/j/shicyoson18/6houkoku/houkoku.pdf. Accessed March 20, 2021.

National Institute of Population and Social Security Research. (n.d.) *Estimated future population and number of households*. Retrieved from http://www.ipss.go.jp/syoushika/tohkei/MainmMai.asp. Accessed March 1, 2021.

National Land Policy Bureau, Ministry of Land, Infrastructure, Transport and Tourism, C. (2015). *Method for estimating the future estimated population by 500m and 1km mesh based on the 2015 census*. Nlftp.mlit.go.jp. Retrieved from https://nlftp.mlit.go.jp/ksj/gml/datalist/mesh500_1000_h30.pdf. Accessed March 1, 2021.

National Statistics Center. (2021). *e-Stat Portal Site of Official Statistics of Japan*. Retrieved from https://www.e-stat.go.jp. Accessed March 1, 2021.

Rubinyi, S., Blankespoor, B. & Hall, J. (2021). The utility of built environment geospatial data for high-resolution dasymetric global population modeling. *Computers, Environment and Urban Systems*, 86, 101594.

Sinha, P., Gaughan, A. E., Stevens, F. R., Nieves, J. J., Sorichetta, A. & Tatem, A. J. (2019). Assessing the spatial sensitivity of a random forest model: Application in gridded population modeling. *Computers, Environment and Urban Systems*, 75, 132-145.

Sugimoto, T., Kaminaga, N., Kato, S., Takamori, S., & Sato, T. (2018). A practical location equilibrium model to study measures promoting compact city policy. *Journal of Japan Society of Civil Engineers*, Ser. D3 (Infrastructure Planning and Management), 74(5), I_439–I_451.

Tamura, S., Iwamoto, S. & Tanaka, T. (2018). The impact of spatial population distribution patterns on $CO_2$ emissions and infrastructure costs in a small Japanese town. *Sustainable Cities and Society*, 40, 513–523.

Ishigami, T. & Kurokawa, T. (2001). A Study on population projection method in new towns. *Journal of the City Planning Institute of Japan,* 36, 463–468.

Tsuboi, S., Ikaruga, S. & Kobayashi, T. (2016). Method for the proposal and evaluation of urban structures for compact cities using an expert system. *Frontiers of Architectural Research*, 5(4), 403–411.

Waddell, P. (2002). UrbanSim: modeling urban development for land use, transportation, and environmental planning. *Journal of the American Planning Association*, 68(3), 297–314.

Wilson, T.D. (2013). The sequential propensity household projection model. *Demographic Research*, 28, 681–712.

Yamagiwa, K., Fujii, H. & Yoshimura, S. (2017). Multi-agent-based household transition simulation using mesoscopic model. *Transactions of the Japanese Society for Artificial Intelligence*, 32(5), AG16–A_1–10.

Zhou, M., Li, J., Basu, R. & Ferreira, J. (2022). Creating spatially-detailed heterogeneous synthetic populations for agent-based microsimulation. *Computers, Environment and Urban Systems*, 91, 101717.

Zhuge, C., Shao, C., Wang, S. & Hu, Y. (2018). An agent- and GIS-based virtual city creator: A case study of Beijing, China. *The Journal of Transport and Land Use*, 11(1), 1231–1256.

**Footnotes**

1. A method for estimating the future population by multiplying the change rate of births, deaths, migration in a cohort of males and females in five-year age groups; this method is called the "Three Components of Population Change."

2. A method for estimating the number of households by setting the probability that a cohort of men and women in five-year age groups will be the householder in each family type.

3. The formula for calculating the average absolute error rate is as follows:

$$MAPE = \frac{100}{n} \sum_{i=1}^{n} \frac{|ai - fi|}{ai}$$

where *ai* is the measured value and *fi* is the estimated value.